\documentclass[10pt]{article}
\usepackage{graphicx}
\usepackage{amssymb}
\usepackage{amsmath}
\usepackage{color}
\def\be{\begin{equation}}
\def\ee{\end{equation}}
\def\bea{\begin{eqnarray}}
\def\eea{\end{eqnarray}}

\def\<{\langle}
\def\>{\rangle}
\def\~{\tilde}
\def\s{\sigma}

\numberwithin{equation}{section}
\numberwithin{figure}{section}
\numberwithin{theorem}{section}
\numberwithin{teorema}{section}

\def\be{\begin{equation}}
\def\ee{\end{equation}}
\def\bc{\begin{center}}
\def\ec{\end{center}}

\title{\bf A statistical mechanics approach to Granovetter theory}
\author{Adriano Barra\ $^1$, Elena Agliari\ $^2$}
\begin{document}
\date{}
\maketitle

\begin{center}
{\small
\vskip-0.5cm


\footnote{e-mail:{\tt  adriano.barra@roma1.infn.it}} Dipartimento
di Fisica, Sapienza Universit\`a di Roma (Italy) \vskip-0.5cm Gruppo
Nazionale per la Fisica Matematica, Sezione di Roma1 \vskip-0.5cm
\footnote{e-mail:{\tt elena.agliari@fis.unipr.it}} Dipartimento di
Fisica, Universit\`a di Parma (Italy) \vskip-0.5cm Istituto Nazionale di
Fisica Nucleare, Gruppo Collegato di Parma  \vskip-0.5cm Theoretische Polymerphysik, Freiburg Universit\"at Freiburg (Germany)}
\end{center}


{\small \bf------------------------------------------------------------------------------------------------------------}

\vskip-1cm
{\small

{\bf Abstract.}

In this paper we try to bridge breakthroughs in quantitative
sociology/econometrics  pioneered during the last decades by Mac
Fadden, Brock-Durlauf, Granovetter and Watts-Strogats through
 introducing a minimal model able to reproduce essentially all the
features of social behavior highlighted by these authors.
\newline
Our model relies on a pairwise Hamiltonian for decision maker
interactions which naturally extends the multi-populations
approaches by shifting and biasing the pattern definitions of an
Hopfield model of neural networks. Once introduced, the model  is
investigated trough graph theory (to recover Granovetter and
Watts-Strogats results) and statistical mechanics (to recover
Mac-Fadden and Brock-Durlauf results). Due to internal symmetries
of our model, the latter is obtained as the relaxation of a proper
Markov process, allowing even to study its out of equilibrium
properties.
\newline
The method used to solve its equilibrium is an adaptation of the
Hamilton-Jacobi technique recently introduced by Guerra in the
spin glass scenario and the picture obtained is the following:
just by assuming that the larger the amount of similarities among
decision makers, the stronger their relative influence, this is
enough to explain both the different role of strong and weak ties
in the social network as well as its small world properties. As a
result,  imitative interaction strengths seem essentially a robust
request (enough to break the gauge symmetry in the couplings),
furthermore, this naturally leads to a discrete choice
modelization when dealing with the external influences and to
imitative behavior a la Curie-Weiss as the one introduced by Brock
and Durlauf.
%
%
}


\cleardoublepage


\section{Summarizing some main results of quantitative sociology}

In recent years there has been an increasing awareness towards the
problem of finding a quantitative way to study the role played by
human interactions in shaping behavior observed at a population
level, ranging from the context of pure sociology to the one
belonging to economic sciences. The conclusion reached by all
these studies is that mathematical models have the potential of
describing several features of social behavior, among which, for
example, the sudden shifts often observed in society's aggregate
behavior \cite{nesh}\cite{bouchaud2}\cite{kuran}, and that these
are unavoidably linked to the way individual people influence each
other when deciding how to behave (the phase transitions in the
language of thermodynamics \cite{ellis}), the whole suggesting a
promising potential application of disordered statistical
mechanics to this field of research
\cite{capitolo}\cite{bucca}\cite{durlauf2}.
\newline
Here we summarize what we understood as real breakthroughs in
these analysis, highlighting two main aspects dealing with
topological investigations on the structure of the graph built by
social interactions and the kind of interactions themselves.
Namely, the discovery of the fundamental role of weak ties in
bridging different communities (due to Granovetter
 \cite{bucca,grano1,grano2,grano3}) and the ``small world'' feature of the
social structure (obtained by Watts and Strogatz
\cite{callaway,watts,watts2}) for the first analysis and the
discrete choice of decision makers in econometric (due to Mac
Fadden \cite{barPLgallo,mcfadden,follmer}) and the essentially
imitative behavior among these agents (due to Brock and Durlauf
\cite{durlauf,durlauf2,ellis}).
%
%

Even though fundamental experiments dealing with social networks
may constellate modern society analysis (i.e. the paradigmatic
Milgram experiment of the sixties \cite{milgram}), a real
breakthrough in our understanding of network structure inside
modern societies has been achieved when Granovetter reversed the
Chicago school of social-psychology showing how a person with
built weak ties (which were previously seen, at individual level,
as ancestors of depressive states) was much more able to adapt its
behavior to the social fitness due to the much broader amount
of available information: in particular he noticed that these weak
ties may often carry information of little significance (but not
redundant as in highly clusterized community of similar agents
linked by strong ties), however they allow a primarily
transmission of new information across otherwise disconnected
cluster of the social network (allowing great potential benefit by
these bridges).
%
%
\newline
Two decades after this achievement, Watts and Strogatz, trough a
mathematical technique (rewiring) have been able to display the
Milgram results (sometimes known as ``six degrees of separation")
by which they understood that social (as well as others, i.e.
biological \cite{AB}\cite{barabasi}) networks can not be described
by purely ordered or purely random graphs (i.e. Erdos-Renyi ones
\cite{barabasi,newman}) due to correlations among nodes which
allow for a much faster transmission of information (real graphs
show high degree of cliqueness \cite{new_lett,vespignani}),
pioneering a quantitative approach to these new networks, nowadays
often called "small worlds".
%

In a different but related context, Mac Fadden has shown how to
infer a model for econometric estimation of binary decision making
(reflecting accurately several real cases in social structures
\cite{fadden}) by introducing fundamental dichotomic degrees of
freedom inside each agent mirroring its personal attitudes (i.e. a
bit string $\bold{\xi}$ of $K$ entries $\mu=1,...,K$ where each
entry represent an attribute, i.e. $\mu=1$ accounting for smoking
such that $\xi_i^{\mu=1}=+1$ states that the i agent smokes while
$\xi_i^{\mu=1}=0$  states that he does not, and so on). Once
indexed individuals by $i, \ i=1...N$, and assigned an Ising spin
to each individual's choice $\s_i=+1$ for the agreement or
$\s_i=-1$ for disagreement, he chooses to exploit data by assuming
a single particle model into a suitable external field $h_i$ (the
``field'' influencing the choice of $i$) which is a function of
the vector of attributes $\xi_i$. Since for the sake of simplicity
  attributes are taken as binary variables, the whole theory can be
described in terms of an effective one-body Hamiltonan $H$ as
$$
H_N(\s;\xi) =\sum_i^N h_i \sigma_i = \frac{h}{K} \sum_i^N
\sum_{\mu}^K\xi_h^{\mu}\xi_i^{\mu}\sigma_i,
$$
where $h$ is a scalar parameter ruling the overall intensity of
the external stimulus (whose capabilities of influencing a generic
agent $i$ are encoded into its bit-string $\xi_h$). This
parametrization of $h_i$ correspond to what economists call a
\emph{discrete choice} model \cite{mcfadden}, and shows a
remarkable link between econometrics and statistical mechanics
($H_N(\s;\xi)$ can be seen as a random field Ising model): In fact
discrete choice theory has the same variational flavour of
thermodynamics as states that, when making a choice, each person
weights out various factors such as his own gender, age, income,
etc, as to maximize in probability the benefit arising from
his/her decision.


Despite this result, there exist many examples from  economics and
sociology where it has been observed how the global behavior of
large groups of people can change in an abrupt manner as a
consequence of slight variations in the social structure (such as,
for instance, a change in the pronunciation of a language due to a
little immigration rate, or as a substantial decrease in crime
rates due to seemingly minor action taken by the authorities
\cite{kuran,critmass}). From a statistical mechanical point of
view, these abrupt transitions should be considered as phase
transitions caused by the interaction between individuals that can
not be accounted by a pure one-body theory. Indeed, Brock and
Durlauf have shown \cite{durlauf} how discrete choice can be
extended to the case where a global mean-field interaction is
present (providing an interesting mapping to the Curie-Weiss
theory (CW) \cite{ellis, barraJSP}), thus further highlighting the
close relation existing between the econometric and the
statistical mechanical approaches to these problems.

Instead of introducing Brock-Durlauf approach (which is a
systematic translation of the CW scenario in social sciences) we
go one step forward following the subsequent generalization
obtained by diving the ensemble of the N decision makers into
clusters, due to Contucci and coworkers \cite{barPLgallo,
PLgallo}: Introducing a general two-body Hamiltonian $H_N(\s;J)$
as
\begin{equation}\label{ham}
H_N(\s;J)=-\sum_{i,l=1}^N J_{il}\s_i\s_l - \sum_{i=1}^N h_i \s_i,
\end{equation}
they went over by defining  a suitable parametrization for the
interaction coefficients $J_{il}$. Since each agent is
characterized by $k$ binary socio-economic attributes, the
population can be naturally partitioned into $2^k$ subgroups,
which for convenience are taken of equal size: this leads to
consider a mean field kind of interaction, where coefficients
$J_{il}$ depend explicitly on such a partition as follows
\begin{eqnarray*}
J_{il}=\frac{J}{2^k \, N}\delta_{gg'}, \ \textrm{if $i \in g$ and
$l \in g'$},
\end{eqnarray*}
which in turn allows us to rewrite (\ref{ham}) as
\begin{eqnarray*}\label{intens}
H_N(\s;J)=-\frac{NJ}{2^k}(\sum_{g,g'=1}^{2^k} \delta_{gg'}m_g
m_{g'} + \sum_{g=1}^{2^k} h_g m_g)
\end{eqnarray*}
where $m_g$ is the average opinion of group $g$, namely $
m_g=\frac{1}{2^k \, N}\sum_{i=(g-1)N/2^k+1}^{g \, N/2^k} \s_i$.
\newline
This idea  of partitioning the interaction matrix into clusters of
similar agents can be extended to a natural limit (that we work
out here) such that the size of these clusters approaches zero in
the thermodynamic limit (so to preserve each individual identity
and uniqueness for each agent, or a measure of attitude
fluctuations inside the original idea of clusters): interestingly
this leads to an interaction matrix so that \be\label{hebb} J_{ij}
= \frac1K \sum_{\mu}^K \xi_i^{\mu}\xi_j^{\mu}, \ee and naturally
collapses the concept of magnetization in spin glasses \cite{MPV}
to the one of retrieval in neural networks \cite{amit}, ultimately
switching frustration into dilution (as $\xi \in [0,+1]$ instead
of $\xi \in [-1,+1]$).

\section{The model and its topology: graph theory}

In this section we look at the population as a graph and we study
its topological properties: each agent $i$ is represented by a
node; couples of agents $(i,j)$ displaying a positive coupling
$J_{ij}>0$ are said to be in contact or to interact with each
other and this is envisaged by means of a link between $i$ and
$j$, whose weight is just $J_{ij}$ (cft. eq.\ref{hebb}). This
picture mirrors the idea that socio-economic relations between
individuals or firms are embedded and organized in actual social
networks, which follows from the seminal work by Granovetter.

As anticipated, each agent $i$ is characterized by a binary string
$\xi_i$, which might be thought of as the codification of agent's
attitude, either positive ($1$) or negative ($0$), towards a given
issue. All strings are taken of length $K$ and each entry is
extracted randomly according to
\begin{equation} \label{eq:character_distribution}
P(\xi_i^{\mu}=+1)=  \frac{1+a}{2}, \;\; P(\xi_i^{\mu}= 0)=
\frac{1-a}{2},
\end{equation}
in such a way that, by tuning the parameter $a \in [-1,+1]$, the
concentration of non null-entries for the $i$-th string $\rho_i =
\sum_{\mu} \xi_i^{\mu}$ can be varied; consistently, also the
topology generated by the rule in Eq.~\ref{hebb} is varied. In
particular, when $a \to -1$ the system is completely disconnected
(and only discrete choice survives), while when $a \to +1$ each
link is present, being
$J_{ij}=1$ for any couple (so to retain the fully Brock-Durlauf
approach). As we will see, small values of $a$ give rise to highly
correlated, diluted networks, while, as $a$ gets larger the
network gets more and more connected and correlation among links
vanishes. In agreement with our modelization intent, repetitions
among strings are allowed.

The main topological features of the emergent network have been
investigated in \cite{AB,BA}, where it was shown that the average
link probability among two generic nodes is
\begin{equation} \label{eq:p}
p = 1 - \left[ 1- \left( \frac{1+a}{2}\right)^2 \right]^K,
\end{equation}
and that, for large enough $N$ and $K$, with $K$ growing slower
than linearly with $N$, the degree distribution is multimodal.
Therefore, the average degree for a generic node reads as $z = p
N$. \newline Apart from these global, long-scale features, the
model also displays interesting properties concerning correlation
among links, as we are going to deepen.

Small-world (SW) networks are characterized by two basic
properties, that are a large clustering coefficient, i.e. they
display subnetworks where almost any two nodes within them are connected, and a small
diameter, i.e. the mean-shortest path length among two nodes grows
logarithmical with $N$. While the latter requirement is a common
property of random graphs \cite{newman,barabasi}, and is therefore
satisfied also by the graph under study, the clustering
coefficient deserves more attention.
Several attempts in the past have been made in order to define
network models able to display such a feature
\cite{newman,barabasi,watts}. For instance, in their seminal work,
Watts and Strogatz \cite{watts} introduced a rewiring procedure on
links, which can yield the desired degree of correlation. As we
are going to show, in our approach SW effects emerge naturally
from the definition of patterns and from the rule in
Eq.~\ref{hebb}, that is to say, interactions based on sharing of
interests (i.e. non-null entries) intrinsically generate a
clustered society.

\begin{figure}[tb]
\includegraphics[height=100mm]{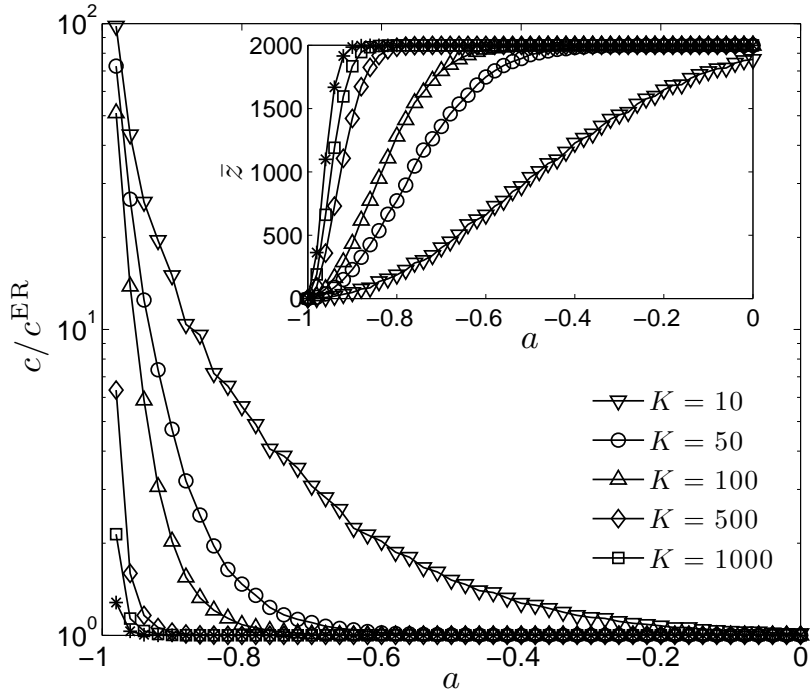}
\caption{\label{fig:clustering} Main figure: Ratio between the clustering
coefficient $c$ measured for the correlated network under
consideration and the coefficient $c^{\mathrm{ER}}$ corresponding
to a purely random graph displaying an analogous density of links,
as a function of the parameter $a$ and for different values of
$K$; the volume $N$ is kept fixed and equal to $2000$. Inset: behavior of the average degree $\bar{z}$ with respect to $a$ for the same set of realizations corresponding to the main figure.}
\end{figure}

Before proceeding, we notice that the same property can be
addressed in different ways: we can say that the graph exhibits a
large transitivity, meaning that if $i$ is connected to both $j$
and $k$, then $j$ and $k$ are likely to be connected; in modern
network theory we say that the graph displays a large
``cliquishness'' and we measure it by means of the so-called
clustering coefficient \cite{newman,vespignani}: The local
clustering coefficient for node $i$ is defined as
\begin{equation}
c_i = \frac{2 E_i}{z_i (z_i -1)},
\end{equation}
where $E_i$ is the number of actual links present within the
neighborhood of $i$, whose upper bound is just $z_i(z_i-1)/2$,
namely is the number of connections for a fully connected group of
$z_i$ neighbors nodes. Then, the (global) clustering coefficient
for the whole graph simply follows as the average of $c_i$ over
all nodes. Usually, for a graph with average degree equal to $z$,
having a large clustering means that $c$ is larger than $z/N=p$,
which represents the clustering coefficient $c^{\mathrm{ER}}$ for
an ER random graph with an analogous degree of dilution.
\newline
As for the graph under study, we found that \cite{BA} $c \approx p
+ 1/ \rho - 1/(z-1) >p$, where in the last equality we used $\rho
< z-1$, which holds when $N$ is large enough and the graph
topology non-trivial \cite{BA}. More generally, one can notice
that for this graph, the $z_i$ neighbors of $i$ are all nodes
displaying at least one non-null entry corresponding to any
non-null entries of $\xi_i$; this condition biases the
distribution of strings relevant to neighboring nodes, so that
they are more likely to be connected with each other. Indeed, when
$\rho_i=1$, it is easy to see that $c_i$=1, to be compared with
$c^{\mathrm{ER}}_i$, namely the average link probability for node
$i$, which turns out to be $(1+a)/2 \leq 1$; analogous arguments
apply also for larger values of $\rho$ \cite{BA}. A numerical
corroboration can be found in Fig.~\ref{fig:clustering}, which
shows that $c > c^{\mathrm{ER}} $ in a wide region of values of
$a$ corresponding to non-trivial networks, i.e. for $a$ larger
than the percolation threshold and smaller than the
fully-connected threshold. The clustering effect is especially
manifest  in the region of high dilution, where, for graphs
analyzed here, $c$ is even two orders of magnitude larger than
$c^{\mathrm{ER}}$. Of course, when $a$ approaches $1$, the graph
gets fully connected and $c \to c^{ER} \to 1$.  These results are
robust as $N$ is varied.

\begin{figure}[tb]
\includegraphics[height=100mm]{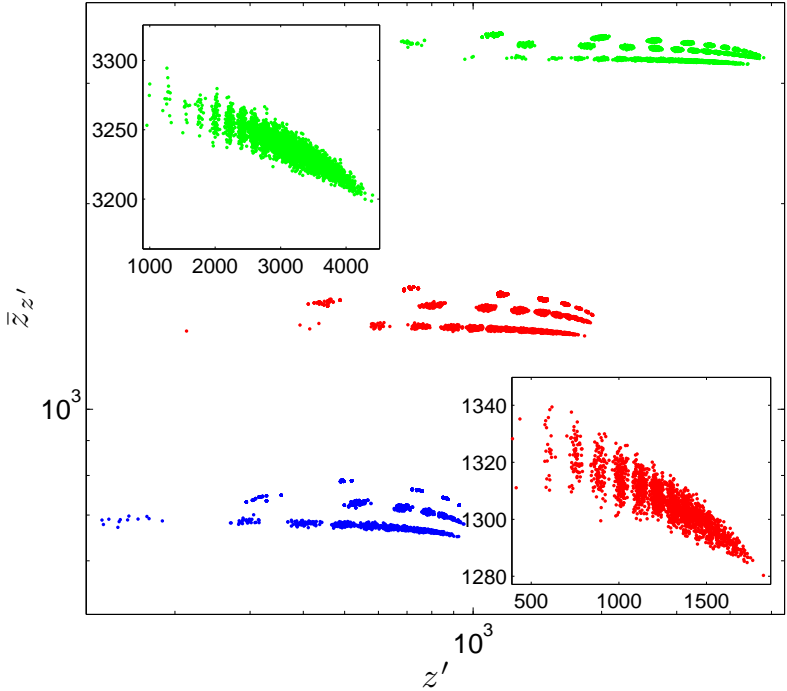}
\caption{\label{fig:assorta1} Neighbor assortativity for different
system sizes (from top to bottom $N=2000, N=5000, N=8000$,
depicted in different colors) and different values of $K/N$ (for
each choice of $N$, from top to bottom, $K/N = 25, K/N=100, K/N =
250$). Notice that for $K$ small different modes can be
distinguished. The two insets show in detail the case $N= 2000,
K=80$ (upper panel) and $N=8000, K=320$ (lower panel). The value
of $a$ is fixed and equal to $-0.6$.}
\end{figure}

Finally, we mention another  quantity used in ecology and
epidemiology to quantify the existence of correlations among
links, that is the so-called \emph{assortativity coefficient}
\cite{newman,new_lett}: a network is said to show ``assortative
mixing'' (``dissortative mixing'') on their degrees whenever
high-degree vertices prefer to attach to other high-degree
(low-degree) vertices. While assortativity is typical of social
networks, dissortativity is often found in technological and
biological networks.
\newline
The assortativity coefficient $r$ can be  defined as a Pearson
coefficient to measure the correlation between the coordination
numbers at either ends of a link; the ER graph corresponds to
$r=0$ \cite{newman}. The measures performed on the graph under
study suggest a dissortative behavior ($r<0$), which is
corroborated by the quantity $\langle z \rangle_{z'}$,
representing the average degree over the nearest-neighbors of a
node with degree $z'$, namely:
\begin{equation}
\bar{z}_{z'} = \sum_{z=0}^{N-1} z P(z; z'),
\end{equation}
where $P(z;z') $ is the conditional probability that a link
stemming from a node with degree $z'$ points  to a node with
degree $z$. As shown in Fig.\ref{fig:assorta1}, a decreasing
behavior of $\bar{z}_{z'}$ is consistent with a dissortative
mixing.
%
%
The reason of this behavior is clear to see: while nodes
corresponding to strings with large $\rho$ can connect to most
other nodes, nodes with small $\rho$ have basically no chance to
connect to other similar strings. This gets more evident for $a$
small, as the concentration of such strings is larger.
Interestingly, as highlighted in \cite{new_lett}, dissortativity
has significant  effects on the resilience (see next section) of
the structure itself: dissortatively mixed networks are less
robust to the deletion of their vertices than assortatively mixed
or neutral networks.

\begin{figure}[tb]
\includegraphics[height=90mm]{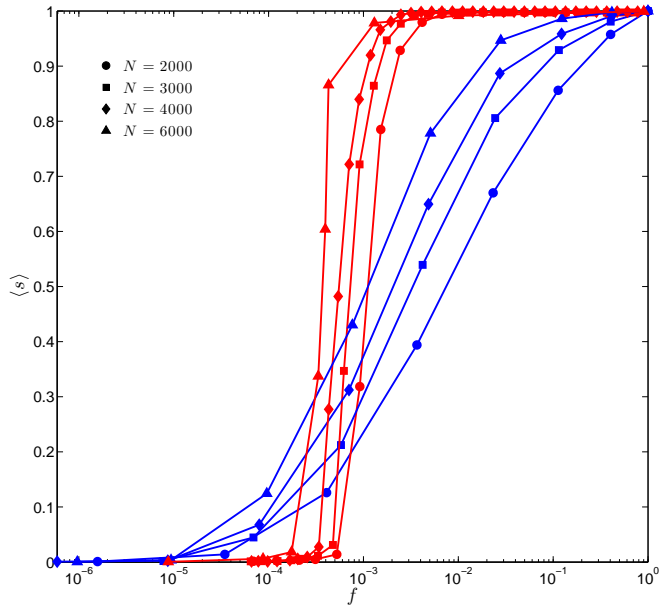}
\caption{\label{fig:generating} \label{fig:weak} Size of the
largest component versus the fraction of links in the system.
Several size are depicted $N=2000$,  $N=3000$, $N=4000$, $N=6000$,
all corresponding to the same value $\alpha=100$ and $\gamma=1$.
Notice that when weak ties are deleted first, the way the size of
the largest component $\langle s \rangle$ increases with $f$ is
smoother with respect to the random delation and also that
$\langle s \rangle$ is smaller than $1$ at relatively large values
of $f$.}
\end{figure}

%
As first remarked in \cite{grano1}, real social networks are not
only characterized by a small-world topology, which basically
means large clustering and small diameter, but they also feature a
peculiar coupling pattern. In fact, not only the neighbors of a
given node $i$ are likely to be connected, but they form
communities such that intra-group links are expected to be
stronger than inter-group links. In this way weaker ties work as
\emph{bridges} connecting communities strongly  linked up. Interestingly,
analogous properties are found also for $\mathcal{G}(N, K, a)$, in
fact, it is intuitive to see that nodes displaying very similar
strings are likely to be intensively connected with each other,
hence forming a group, while, each of them, separately and
according to the pertaining string, can be weakly connected with
other nodes/groups.
\newline
In order to deepen the role of weak ties, we perform two
percolation processes where links are deleted either
deterministically or randomly: Given the pattern of couplings, in
the former case we delete those with magnitude lower than a given
threshold $\iota \in [0,1]$ meant as a tunable parameter; in the
latter case we progressively delete nodes in a random way. We call
$1-f$ the fraction of links erased (in the former case $f$ is a
function of $\iota$) and we measure the size of the largest
connected cluster (see also \cite{guada} for more details).
Results are shown and compared in Fig.~\ref{fig:weak}.
Interestingly , when weak links are deleted first the graph starts
to be disconnected at a value of $f$ rather small, and this is a
signature that such links work as bridge. On the other hand, strong ties are highly redundant \cite{guada}. Indeed,
starting from a connected graph, the first nodes to get
disconnected are those with small $\rho$, to fix ideas, those with
$\rho=1$; among these, the ones with the non-null entry in the
same position were completely clustered in the original network.
As the threshold $\iota$ is increased, more and more nodes get
disconnected; most of them remain isolated (typically those with
$\rho < \iota$), however, some non-trivial components survive.
Such clusters are made up of very similar strings with a
relatively large number of non-null entries ($\rho > \iota$) and
are therefore all closely connected.

It is worth noting that such strongly clustered components emerge just in the ``critical region'', namely where it is possible to detect nodes bridging two clusters and which play as ``brokerage'' between distinct group; this is a strategic position since it allows access to a more diverse set of ideas and information. The notions of homogeneity within groups and intermediacy between groups form the basis for the theory of
"structural holes" introduced by Burt \cite{burt}.



We finally comment on the resilience properties of the network
under study, which can be as well inferred from the analysis on
percolation processes. According to the situation, the stability
of the network can be defined as its ability to remain connected
or to still exhibit a giant component, under edge removal. In the
former case, if weak links are the most prone to failure, our
correlated network performs rather badly. Conversely, if we are interested in the
maintenance of a macroscopic connected component, given that weak
links are the first to be deleted, our correlated network performs
definitely better, as the percolation threshold grows slowly with $N$ (see also \cite{guada}).

\section{The model and its relaxation: stochastic dynamics}

We saw that the general structure of the Hamiltonian obeys
\be\label{model} H_N(\sigma;\xi)= \frac{1}{NK}\sum_{i<j}^{N,N}
\sum_{\mu=1}^K \xi_i^{\mu}\xi_j^{\mu}\sigma_i \sigma_j  + \sum_i^N
h_i \s_i. \ee It is worth recalling that the attributes ($\xi$)
are drawn randomly once for all, and so ar treated as quenched
variables: this does not mean that a particular agent does not
evolve in time changing his attribute distribution, but that,
overall, one agent may switch to another and viceversa as far as
the global attribute distribution is kept constant.
\newline
The state of the system at this time is given by the average of
all its building agents (such that we can introduce a
"magnetization" as their average $m = N^{-1}\sum_i^N \sigma_i$),
each of which evolving time-step by time-step via a suitable
dynamics:
\newline
Following standard disordered statistical mechanics approach
\cite{peter} we introduce the latter accordingly to
\begin{equation}\label{markov}
\sigma_i(t+1) = sign \left( \tanh(\beta \varphi_i(t)) +
\eta_i(t)\right ),
\end{equation}
where $\varphi_i(t)$ is the overall stimulus felt by the $i$-th
agent, given by
\begin{equation}
\varphi_i(t) = N^{-1} \sum_j^N J_{ij} \s_j(t) + h_i(t),
\end{equation}
and the randomness is in the noise implemented via the random
numbers $\eta_i$, uniformly drawn over the set $[-1,+1]$.
 $\beta$ rules the impact of this noise on the state
$\sigma_i(t+1)$, such that for $\beta=\infty$ the process
is completely deterministic while for $\beta=0$ completely random.
\newline
In the sequential dynamics we are introducing, at each
time step $t$, a single agent $l_t$ -randomly chosen among the
$N$- is updated, such that its evolution becomes
\be
P[\sigma_{l_t}(t+1)]=\frac12 \big( 1 +
\sigma_{l_t}(t)\tanh(\beta \varphi_{l_t}(t))\big), \ee whose
deterministic zero-noise limit is immediately recoverable by
sending $\beta \to \infty$.
\newline
If we now look at the probability of the state at a given time
$t+1$, $P_{t+1}(\sigma)$, we get \be\label{fokkerplanck}
P_{t+1}(\sigma) = \frac1N \sum_{i}^{N} \frac{1}{2}(1+
\sigma_i\tanh(\beta \varphi_i (\sigma)))P_t(\sigma) + \frac1N
\sum_{i}^{N} \frac{1}{2}(1+ \sigma_i \tanh(\beta \varphi_i
(F_i\sigma)))P_t(F_i\sigma), \ee where we introduced the $N$
flip-operators $F_i$, $i \in (1,...,N)$, acting on a generic
observable $\phi(\sigma)$, as
\begin{equation}
F_i\Phi(\sigma_1,..., + \sigma_i, ...,
\sigma_N)\ =\Phi(\sigma_1,..., -\sigma_i, ...,
\sigma_N)\end{equation}
such that we can write the evolution of the network as a
Markov process
\begin{eqnarray}
p_{t+1}(m) &=& \sum_{m'}W[m;m']p_t(m'),
\\ W[m;m'] &=& \delta_{m,m'} +
\frac1N \sum_{i=1}^N \Big(
w_i(F_i m)\delta_{m, F m'} - w_i(m)
\delta_{m,m'} \Big), \nonumber
\end{eqnarray}
with the transition rates $w_i(m) = \frac12 [1-
\sigma_i \tanh(\beta \varphi_i)]$.

As the affinity matrix is symmetric, detailed balance ensures that
there exists a stationary solution $P_{\infty}(m)$ such that
(restricting $\tilde{h}_i(t) \to \tilde{h}_i \in \mathbb{R} \
\forall i \in (1,...,N)$)
$$
W[m,m']P_{\infty}(m')= W[m',m]P_{\infty}(m).
$$
This key feature ensures equilibrium, which implies \be\label{MB}
p_{\infty}(\sigma;J,h) \propto
\exp\Big(\frac{\beta}{2NK}\sum_{ij}^N \sum_{\mu}^K
\xi_i^{\mu}\xi_j^{\mu} \s_i \s_j - \beta \sum_i^N h_i \s_i\Big) =
\exp\Big(-\beta H_N(\sigma;\xi)\Big), \ee namely the
Maxwell-Boltzmann distribution \cite{ellis,barraJSP} for the
Hamiltonian (\ref{model}).
\newline
In absence of external stimuli, and skipping here the question
about the needed timescales for "thermalization", the system
reaches an equilibrium that it is possible to work out explicitly
and that reproduces all the features stressed in the first section
(as we are going to show).
\newline
For this detailed balanced system furthermore, the sequential
stochastic process (\ref{markov}) reduces to Glauber dynamics such
that the following simple expression for the transition rates
$W_i$ can be implemented \be W_i(m) = \Big( 1 + \exp(\beta \Delta
H(\sigma_i; \xi)) \Big)^{-1}, \ \  \Delta H(\sigma_i; \xi) = H(F_i
m; \xi) - H(m; \xi). \ee

\section{The model and its equilibrium: statistical mechanics}

In the previous section we showed that, if the affinity matrix is
symmetric (i.e. $J_{ij}=J_{ji}$), so that detailed balance holds,
the stochastic evolution of our social model approaches the
Maxwell-Boltzmann distribution (see eq.(\ref{MB})), which
determines the thermodynamic equilibria.

The latter are obtained by extremizing  the free energy
$A(\beta,a,h) = -\beta f(\beta,a,h)= u(\beta,a,h) -
\beta^{-1}s(\beta,a,h)$ ($u$ being the internal energy and $s$
being the intensive entropy) that, as it is straightforward to
check, corresponds to both maximizing entropy and minimizing
energy (at the given level of noise $\beta$, attribute's bias $a$
and external influences $h$). Furthermore, and this is the key
bridge with stochastic processes, there is a deep relation among
statistical mechanics and their equilibrium measure $P_{\infty}$,
in fact \be P_{\infty}(\sigma; \xi,h) \propto \exp(-\beta
H(\sigma; \xi,h)), \ \ A(\beta,a,h) = -\beta f(\beta,a,h) \equiv
\frac{-1}{N} \mathbb{E}\log \sum_{\sigma} \exp(-\beta
H_N(\sigma;\xi,h)). \nonumber \ee The operator $\mathbb{E}$ that
averages over the quenched distribution of attributes $\xi$ makes
the theory not "sample-dependent": For sure each realization of
the network will be different with respect to some other in its
details, but we expect that, after sufficient long sampling, the
averages and variances of observable become unaffected by the
details of the quenched variables.
\newline
Hence, once the microscopic interaction laws are encoded into the
Hamiltonian, we can achieve a specific expression for the free
energy, from which we can derive both the internal energy
$u(\beta,a,h)$ as well as its related entropy $s(\beta,a,h)$:
\begin{eqnarray} u(\beta,a,h) &=& -\partial_{\beta} (\beta
f(\beta,a,h)) = N^{-1}\langle H(\sigma; \xi,h) \rangle, \\
s(\beta,a,h) &=& f(\beta,a,h) + \beta^{-1}\partial_{\beta}(\beta
f(\beta,a,h)).
\end{eqnarray}
The Boltzmann state is given by
\begin{equation}
\omega(\Phi(\sigma,\xi)) = \frac{1}{Z_{N}(\beta,a,h)}
\sum_{\{\sigma_N\}} \Phi(\sigma;\xi) e^{-\beta H_{N}(\sigma,\xi)},
\end{equation}
where the normalization $Z$ is called "partition function" and the
total average $\langle \Phi \rangle$ is defined as
\begin{equation}
\langle \Phi \rangle = \textbf{E}[\omega(\Phi(\sigma,\xi))].
\end{equation}

We want to tackle the problem of solving the thermodynamics of the
model trough the Hamilton-Jacobi technique \cite{sumrule}
\cite{barraJSP}\cite{BGDB}\cite{genovese}.
\newline
Before outlining the strategy, some further definitions are in
order here to lighten the notation (see also \cite{BA} for more details): taken $g$ as a generic
function of the quenched variables we have \be
\mathbb{E}_{\xi}g(\xi) =
\sum_{l_b=0}^{N}\sum_{l_c=0}^{K}\binom{N}{l_b}\binom{K}{l_c}
\left(\frac{1+a}{2}\right)^{l_b+l_c}\left(\frac{1-a}{2}\right)^{N+K-l_b-l_c}\delta_{l_bl_c=l} g(\xi),
\ee
where we summed over the probability $P(l)$ that in the  graph a
number $l$ of weighted links out of the possible $N \times K$
display a non-null coupling, i.e. $ \xi \neq 0 $;  this problem
has been rewritten in terms of $P(l_b)$ and $P(l_c)$, where
$P(l_b)$ is the probability that $l_b$ (out of $N$ random links)
are active and analogously, mutatis mutandis, for $P(l_c)$ (on $K$
random attributes): In fact, $\xi_{i}^{\mu}$ can be looked at as an
$N \times K$ matrix generated by the product of two given vectors
like $\eta$ and $\chi$, namely $\xi_{i}^{\mu}=\eta_i \chi_{\mu}$, in
such a way that the number of non-null entries in the overall
matrix $\mathbf{\xi}$ is just given by the number of non-null
entries displayed by $\eta$ times the number of non-null entries
displayed by $\chi$. Hence, $P(l)$ is the product of $P(l_b)$ and
$P(l_c)$ conditional to $l_b l_c =l$.
\newline
We can  introduce now the following order parameters \be M_{l}=
\frac1N \sum_i^N \omega_{l+1}(\sigma_i), \ee and the Boltzmann
states $\omega_l$ are defined by taking into account only $l$
terms among the elements of the whole involved.
\newline
Namely, $\omega_{l+1}$ has only $l+1$ terms of the type $ \sigma
\sigma$ in the Maxwell-Boltzmann exponential, all the others being
zero: By these ``partial Boltzmann states" we can define the
average of the order parameters as \be\label{parametrodordine}
\langle M \rangle = \sum_{l}^{N-1}P(l)M_{l}. \ee
We are now ready to show our strategy by defining the following
interpolating free energy, depending by two interpolants, $t,x$,
which can be though of as time and space in a mechanical analogy
\cite{sumrule}\cite{barraJSP}\cite{BGDB}\cite{genovese} \be A(t,x)
= \frac1N \mathbb{E}\log \sum_{\sigma}\exp\Big(
\frac{-t}{2NK}\sum_{ij}^N\sum_{\mu}^K\xi_i^{\mu}\xi_j^{\mu}\sigma_i
\sigma_j +  \sum_i^N h_i \sigma_i + x \sum_i^N \tilde{\xi}_i
\sigma_i \Big), \ee where the random links $\tilde{\xi}_i$ have
the same distributions of the standard $\xi_i^{\mu}$ as in any
standard stochastic stability approach. Of course statistical
mechanics is obtained when evaluating this trial free energy at
$t=-\beta, x=0$. Let us work out the derivatives now: \be
\frac{\partial A(t,x)}{dt} = -\frac{1}{2}(\frac{1+a}{2})^2\langle
M \rangle, \ \ \frac{\partial A(t,x)}{dx} = (\frac{1+a}{2})\langle
M \rangle. \ee If we now introduce the following potential
$V(t,x)$: \be V(t,x)=\frac12(\frac{1+a}{2})^2\Big( \langle M^2
\rangle - \langle M \rangle^2 \Big), \ee we can write the
following Hamilton-Jacobi equation for the trial free energy \be
\partial_t A(t,x) + \frac12 \Big(\partial_x A(t,x)\Big)^2 +
V(t,x)=0. \ee When interesting at the replica symmetric regime (in
a nutshell an approximation -which is widely believed to be
correct even though not yet rigorously proved- in which we do not
consider fluctuations of the order parameters in the large size of
the population limit) we simply have to solve the free motion
because replica symmetry means $\lim_{N\to \infty}V(t,x)=0$. The
free field solution is given by the action in a generic point of
the space-time plus the time-integral of the Lagrangian
$\mathcal{L} = (\frac{1+a}{2})^2\langle M^2 \rangle/2$. Namely we
can write \be A(x,t) = A(x_0,0) + \int_0^t \mathcal{L}(t')dt'. \ee
So we have \be A(x_0,0)=\log 2 + \langle \log \cosh (x_0 \xi +
\beta h) \rangle = \frac{1+a}{2}\log\cosh(x_0+\beta h) +
\frac{1-a}{2}\log\cosh(\beta h). \ee The time integral of the
Lagrangian (as there is no potential) is simply
$\frac{1}{2}(\frac{1+a}{2})^2\langle M^2 \rangle t$ and the
equation of motion is a straight line $x(t)=x_0 + \frac{1+a}{2}
\langle M \rangle t$ such that overall we can write at $t=-\beta$
and $x=0 \to x_0 = \frac{1+a}{2}\beta$ \be\label{HJF} A(\beta) =
\log 2 + (\frac{1+a}{2})\log\cosh\Big(\beta[(\frac{1+a}{2}\langle
M \rangle +h)]\Big) + (\frac{1-a}{2})\log\cosh\Big(\beta h\Big)
-\frac{\beta}{2}(\frac{1+a}{2})^2 \langle M \rangle^2.\ee

Now we want to deepen the information encoded into equation
(\ref{HJF}); namely we want to recover by this solution all the
theories of interaction introduced in section one in a
quantitative way.
\newline
Let us start forgetting the network, so with the two limits of
MacFaddend independent particle model $(a \to -1)$ and the pure
Brock and Durlauf theory $(a \to +1)$:
\begin{eqnarray}
\lim_{a\to-1}A(\beta,a) &=& \log2 + \langle \log\cosh(\beta h) \rangle,\\
\lim_{a\to+1}A(\beta,a) &=& \log 2 + \langle \log\cosh\Big(\beta
M_{CW}+h)\Big)\rangle - \frac{\beta}{2}\langle M_{CW}^2 \rangle,
\end{eqnarray}
in perfect agreement with thermodynamics \cite{barraJSP,ellis}.
\newline
Note that when extremizing the free energy with respect to the
order parameter (which is just one in both cases because in the
former, as there is no network, the only decomposition trough eq.
\ref{parametrodordine} is the independent sum of all the
disconnected agents, while in the latter only one graph survives
-the unweighted fully connected- and $P(\langle M \rangle) =
\delta(M - M_{CW})$, where with $M_{CW}$ we meant the standard CW
magnetization), the response of the system is described by the
hyperbolic tangent (nothing but the logit fit function used in
econometrics):
\begin{eqnarray}
\partial_{\langle M \rangle} A(\beta,a=-1,h) &=& 0 \Rightarrow m = \langle \tanh[\beta h]\rangle, \\
\partial_{\langle M \rangle} A(\beta,a=+1,h) &=& 0 \Rightarrow M_{CW} = \langle \tanh[\beta
(M_{CW}+h)]\rangle.
\end{eqnarray}
In all the other cases of interest (so for $a \neq \pm 1$) a distribution for weights on links is always present and weights are stronger for
links among nodes that share higher amount of attribute similarity
as shown in the graph theory analysis.

\bigskip

Last step now should be achieving the critical line, i.e. by the
control of the fluctuations of $\mathcal{M}=\sqrt{N}(M-\bar{M})$.
To fulfil this task it is straightforward to follow the approach
of \cite{AB,BA} (section four), with the streaming now given by
the transport derivative $D =
\partial_t + (\frac{1+a}{2}) \langle \bar{M} \rangle \partial_x$
\cite{sumrule}.  Instead of performing these calculations which
mirrors the ones detailed exposed in the paper \cite{BA} and
depict a phase transition at $\beta_c = (\frac{(1+a)}{2})^{-2}$
(as naively expected), we find instructive to bridge the two
solutions (at $h=0$), namely the one obtained in section tree of
\cite{BA} (eq. $(4.25)$ in that paper) and (\ref{HJF}).
\newline
Starting from the former, let us at first use the self consistency
relation ( eq. $4.24$ of the cited paper) to transform
$\tanh^{-1}(\sum_{l}P(l)\bar{M}_{l})=(\frac{1+a}{2})\beta \langle
\bar{M} \rangle$. This gives
$$
A(\beta,a) = \log 2 + (\frac{1+a}{2})\log\cosh(\beta
(\frac{1+a}{2})\langle \bar{M} \rangle) + \frac{\beta
(\frac{1+a}{2})^2}{2}\langle \bar{M}^2 \rangle - (\frac{1+a}{2})^2
\beta \langle \bar{M} \rangle^2.
$$
The latter can be written exactly as $$A(\beta,a)= \log 2 +
(\frac{1+a}{2})\log\cosh(\beta (\frac{1+a}{2}) \langle \bar{M}
\rangle) -\frac{\beta (\frac{1+a}{2})^2}{2}\langle \bar{M}^2
\rangle$$ by assuming $\langle \bar{M} \rangle^2 = \langle
\bar{M}^2 \rangle$, which, in a nutshell, is sharply the request
$S=0$ (zero source limit) in the double stochastic stability
approach and $V=0$ in the Hamilton-Jacobi approach.

\section{Summary and outlooks}

In this paper we tried to bridge over different aspects of modern
quantitative sociology in a unifying perspective ultimately
offered by a simple shift of the patterns in an Hopfield model of
neural networks. The fundamental prescriptions of the Granovetter
and Watts-Strogatz theories from topological viewpoint and Mac
Fadden and Brock-Durlauf ones from social influences are found as
different limits of this larger model, where, in proper (wide)
regions of the parameters $(\beta,a,h)$, all these features can be
retained contemporarily, offering a systemic view of social
interaction.
\newline
The idea that a model for the associative memory of the brain
(though of as an ensemble of many interacting elementary agents)
may work even for quantifying social behavior is in general
agreement with the "universality" found in all these complex
systems, however, while the neural networks share both positive
and negative links (so to preserve a low synaptic activity, i.e.
$\sum_{\mu}^K \xi^{\mu} \to \mathcal{N}(0,1)$), this property is
avoided in our context (for otherwise the role of weak ties could
be played by highly conflicting peoples). As a consequence,
despite the role of anti-imitative coupling is fundamental (as
discussed for instance in \cite{epl}), it turns out that the
greatest part of social interactions should be essentially
imitative (as sociologists know well from long time). Moreover, by
focusing on couplings generated from the sharing of common
attributes, a small world structure naturally emerges, and,
consistently with real networks, it is possible to detect strongly
clustered sub-communities.
\newline
So, our main goal when dealing with these techniques, is not
discovering other hidden breakthroughs among pioneering
speculations, but offering a quantitative, predictive model (and
related methods for its solution) by which recover agreement with
data and theories and improve society accordingly to our will.
\newline
In this sense, we think that now a great effort must be achieved
in dealing with the inverse problem and its related data analysis,
on which we plan to investigate soon.

\section*{Acknowledgments}

The authors are pleased to thank Raffaella Burioni, Mario Casartelli, Claudia Cioli, Pierluigi
Contucci, Mark Granovetter, Enore Guadagnini and Francesco Guerra for useful discussions.
\newline
This work is supported by the FIRB grant: $RBFR08EKEV$
\newline
GNFM and INFN are also acknowledged.

\end{document}